\documentclass[10pt]{iopart}%as injournal

\usepackage{epsf}

\usepackage{graphicx}
\usepackage{bm}
\begin{document}

%\preprint{}
\title{Ultracold Neutral Plasmas}
\author{ T. C. Killian, Y. C. Chen,   P. Gupta,  S. Laha,  Y. N. Martinez, P. G. Mickelson, S. B. Nagel, A. D. Saenz, and C. E. Simien}
\address{Rice University, Department of Physics and Astronomy
and Rice Quantum Institute, Houston, Texas, 77005}

\vskip10.pt

\begin{abstract}
Ultracold neutral plasmas  are formed by photoionizing laser-cooled atoms
near the ionization threshold. Through the application of atomic physics techniques
and diagnostics, these experiments stretch the boundaries of traditional neutral
plasma physics. The electron temperature in these plasmas ranges from 1-1000\,K and the
ion temperature is around 1\,K. The density can approach $10^{11}$\,cm$^{-3}$.
Fundamental interest stems from the possibility of creating strongly-coupled
plasmas, but recombination, collective modes, and thermalization in these
systems have also been studied.
Optical absorption
images of a strontium plasma, using the Sr$^+$
${^2S_{1/2}} \rightarrow {^2P_{1/2}}$ transition at
422 nm, depict the density profile of the plasma, and probe kinetics on a
50 ns time-scale. The Doppler-broadened ion absorption spectrum measures the ion
velocity distribution, which gives an accurate measure of the ion dynamics in the first
microsecond after photoionization.

\end{abstract}

\vskip10.pt
\section{Introduction}
The study of
ionized gases in neutral plasmas traditionally spans
temperatures ranging from $10^{16}$\,K in the
magnetosphere of a pulsar to $300$\,K in the earth's ionosphere.
Ultracold neutral plasmas \cite{kkb99}, formed by photoionizing
laser-cooled atoms near the ionization threshold, stretch the
boundaries of these studies. The electron
temperature  ranges from 1-1000K and the ion
temperature is around 1 K. The density can approach $10^{11}$
cm$^{-3}$.

Fundamental interest in these systems stems from a range of
phenomena in the ultracold regime. It is possible to form
strongly-coupled  plasmas \cite{ich82} - systems in which the electrical
interaction energy between the charged particles exceeds the
average kinetic energy. This reverses the traditional energy
hierarchy that underlies our normal understanding of plasmas based
on concepts such as Debye screening and hydrodynamics.
Strongly-coupled plasmas appear in exotic environments, such as
dense astrophysical systems \cite{vho91},  matter irradiated with
intense-laser fields \cite{nmg98, sht00}, dusty plasmas of highly
charged macroscopic particles \cite{mtk99}, or non-neutral trapped
ion plasmas \cite{mbh99} that are
 laser-cooled until they freeze into a Wigner crystal. Ultracold
 plasmas are excellent systems in which to study strong-coupling
 achieved at relatively low-density in neutral systems where
 recombination, collective modes, and thermalization are all
interesting things to explore.

Early experiments with ultracold neutral plasmas
used charged particle detection techniques
to study
methods and conditions for forming the
plasma\cite{kkb99}, excitation and detection of plasma
oscillations\cite{kkb00}, dynamics of the plasma
expansion\cite{kkb00}, and collisional recombination into Rydberg
atomic states\cite{klk01}. In related
experiments, researchers at the University of Virginia and
Laboratoire Aim\'{e} Cotton, Orsay, France studied the spontaneous
evolution of a dense, cold cloud of Rydberg atoms into a plasma
\cite{rtn00}.

Optical absorption imaging and
spectroscopy, demonstrated \cite{scg04} using the Sr$^+$
${^2S_{1/2}} \rightarrow {^2P_{1/2}}$ transition in a strontium
plasma, opens many new possibilities. Images depict the density
profile of the plasma, and the Doppler-broadened absorption
spectrum measures the ion velocity distribution. Both can probe
 ion dynamics with 50\,ns resolution.
This paper will concentrate on the physics we have learned from plasma imaging.
A review of the results from studies using charged particle diagnostics can
be found in \cite{kag03}.

Section \ref{Overview} provides an overview of the creation of an ultracold
neutral plasma
and the absorption imaging technique. Section \ref{plasmadynamics}
discusses the dynamics of the ions during the first few
microseconds after photoionization.
Section \ref{spectrumsection} describes the ion absorption spectrum, and
 Sec.\ \ref{resultssection} describes recent studies of plasma equilibration.

%, and for Sec.\
%\ref{model}, which presents  simple models for the evolution of
% the observed ion velocity
%distribution. These models allows quantitative discussion of the
%ion equilibration and subsequent expansion.

\section{Experimental Overview}
\label{Overview}

%\begin{figure}
%  % Requires \usepackage{graphicx}
%\includegraphics[width=3in,clip=true,trim=55 370 20 160,angle=0 ]{apparatus.eps}\\
%  \caption{From \cite{scg04}.
%  Experimental schematic for strontium plasma experiments.
%   The MOT for neutral atoms consists of a pair of anti-Helmholtz magnetic
%  coils and 6  laser-cooling beams. Atoms from a Zeeman-slowed
%  atomic beam enter the MOT region and are trapped.
%  {$^1P_1$} atoms are ionized by the photoionizing laser.
%  The imaging beam passes through the plasma and
%falls on a CCD camera.}\label{apparatus}
%\end{figure}

\begin{figure}
 \mbox{\small
 \begin{minipage}[b]{5.15in}
  \begin{minipage}[t]{1in}
        \mbox{}\\ \unitlength0.75cm
           \includegraphics[width=3in,clip=true,trim=55 220 20 90,angle=0 ]{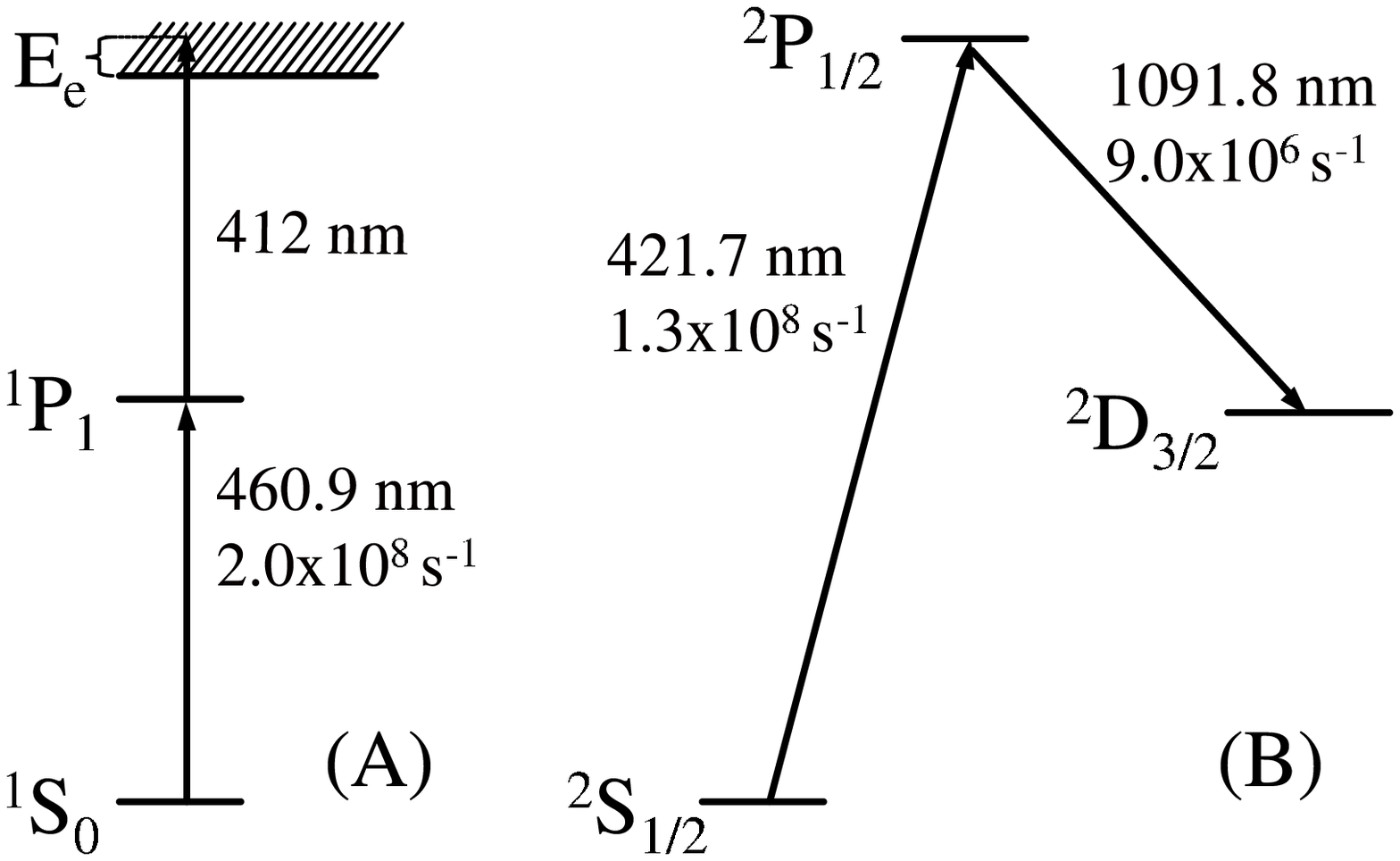}\\
  \end{minipage} \hfill
  \raisebox{.15in}{\parbox[t]{3.1in}{\makebox[0cm]{}{\caption{ Strontium atomic and ionic energy levels involved in the experiment, with decay rates.
  (A) Neutral atoms are laser cooled and trapped in a magneto-optical trap (MOT) operating on the
   ${^1S_0}-{^1P_1}$ transition at 460.9 nm, as
described in \cite{nsl03}. Atoms  excited to the $^1P_1$ level by
the MOT lasers are ionized by photons from a laser at $\sim
412$~nm.
 (B) Ions are imaged using the $^2S_{1/2}-{^2P_{1/2}}$ transition at $421.7$~nm.
 $^2P_{1/2}$ ions decay to the $^2D_{3/2}$ state 7\% of the time, after which
 they cease to interact with the imaging beam.
This does not complicate the experiment because ions typically
scatter fewer than one photon during the time the imaging beam is
on.}\label{energylevels}
}}}
 \end{minipage}}
\end{figure}

\begin{figure}
 \mbox{\small
 \begin{minipage}[b]{5.15in}
  \begin{minipage}[t]{1in}
        \mbox{}\\ \unitlength0.75cm
         \includegraphics[width=3.5in,clip=true,trim=55 370 20 160,angle=0 ]{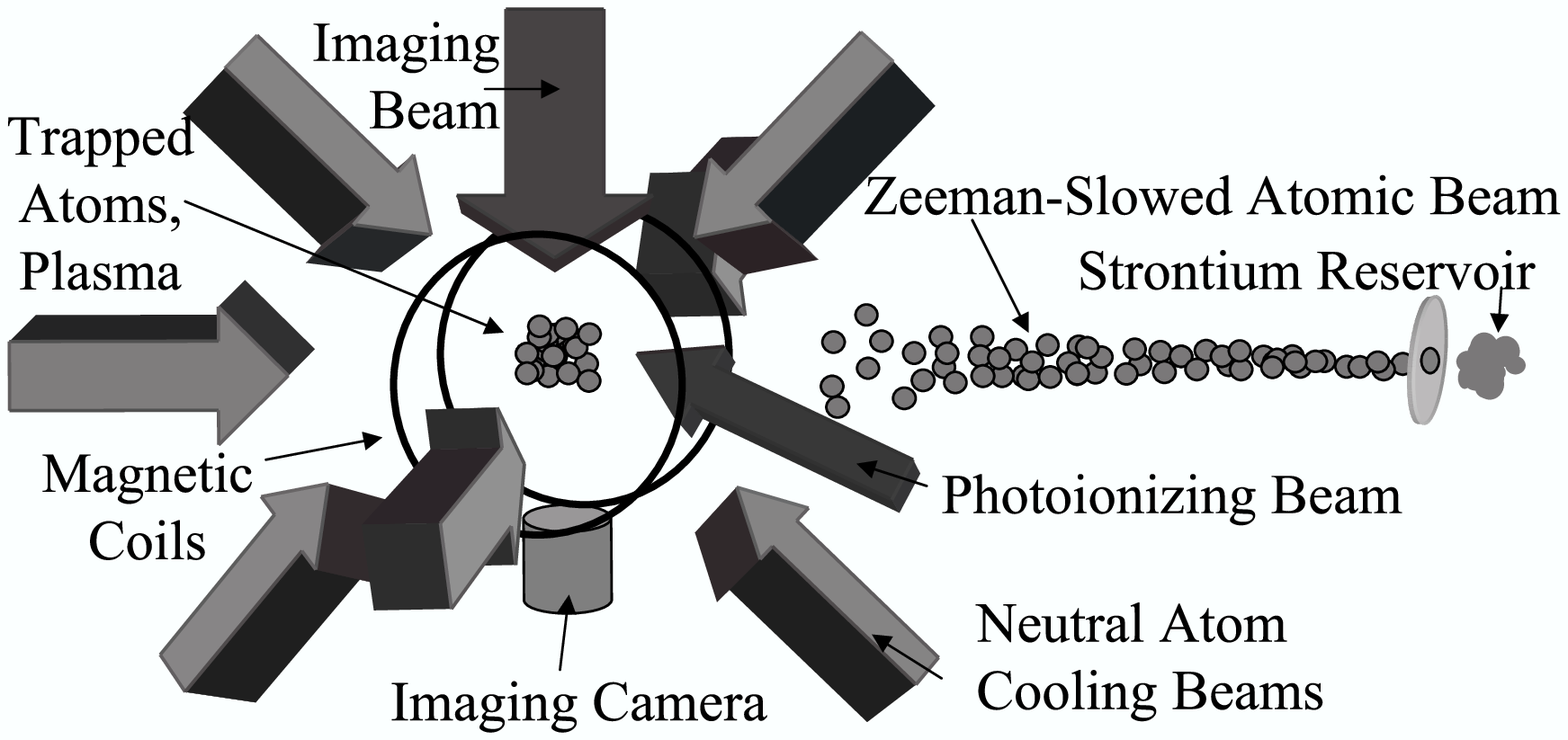}\\
  \end{minipage} \hfill
    \raisebox{.25in}{\parbox[t]{2.6in}{\makebox[0cm]{}{ \caption{From \cite{scg04}.
  Experimental schematic for strontium plasma experiments.
   The MOT for neutral atoms consists of a pair of anti-Helmholtz magnetic
  coils and 6  laser-cooling beams. Atoms from a Zeeman-slowed
  atomic beam enter the MOT region and are trapped.
  {$^1P_1$} atoms are ionized by the photoionizing laser.
  The imaging beam passes through the plasma and
falls on a CCD camera.}\label{apparatus} }}}
 \end{minipage}}
\end{figure}

The recipe for an ultracold neutral plasma starts with
laser-cooled and trapped neutral atoms\cite{mvs99}.
Photons from properly arranged laser beams
scatter off the atoms to generate the forces for cooling and
trapping.  Alkali atoms, alkaline-earth atoms, and metastable
noble gas atoms are the most common atoms for these experiments
because they posses
electric-dipole allowed transitions at convenient laser
wavelengths.
The normal trapping configuration is
called a
magneto-optical trap (MOT) (Fig.\
\ref{energylevels} and \ref{apparatus}) \cite{nsl03}.
%Depending upon the element chosen, up to $10^{10}$
%atoms can be trapped . The density can be as high as $10^{12}
%\,\,{\rm cm}^{-3}$, and the temperature can be in the
% microkelvin to millikelvin  range.
%The production of a strontium plasma starts with atoms that are
%cooled and confined in a magneto-optical trap (MOT) (Fig.\
%\ref{energylevels}). This aspect of the
%experiment was described in \cite{nsl03}.
In the experiments described here, the neutral atom cloud
is characterized by a temperature of a few mK and a density
distribution given by $n({r})=n_0{\rm exp}(-r^2/2\sigma^2)$, with
$\sigma \approx 0.6$~mm and $n_0 \approx 6 \times
10^{10}$\,cm$^{-3}$. The number of trapped atoms is typically $2
\times 10^8$. These parameters can be adjusted. In particular,
turning off the trap and allowing the cloud to expand yields
larger samples with lower densities.

%\begin{figure}
%  % Requires \usepackage{graphicx}
%    \includegraphics[width=3in,clip=true,trim=55 220 20 90,angle=0 ]{energy.eps}
%%\includegraphics[width=3in,clip=true,trim=lefthoriz lowervert rightthoriz uppervert ]
% % \includegraphics[width=4in,clip=true]{energy.eps}\\
%  \caption{ Strontium atomic and ionic energy levels involved in the experiment, with decay rates.
%  (A) Neutral atoms are laser cooled and trapped in a magneto-optical trap (MOT) operating on the
%   ${^1S_0}-{^1P_1}$ transition at 460.9 nm, as
%described in \cite{nsl03}. Atoms  excited to the $^1P_1$ level by
%the MOT lasers are ionized by photons from a laser at $\sim
%412$~nm.
% (B) Ions are imaged using the $^2S_{1/2}-{^2P_{1/2}}$ transition at $421.7$~nm.
% $^2P_{1/2}$ ions decay to the $^2D_{3/2}$ state 7\% of the time, after which
% they cease to interact with the imaging beam.
%This does not complicate the experiment because ions typically
%scatter fewer than one photon during the time the imaging beam is
%on.}\label{energylevels}
%\end{figure}

To form the plasma, the MOT magnets are turned off and atoms are
ionized with photons from the cooling laser and from a $10$~ns
pulsed dye laser whose wavelength is tuned just above the
ionization continuum (Fig.\ \ref{energylevels}). Up to $30$\% of
the neutral atoms are ionized, producing plasmas with a peak
electron and ion density as high as $n_{0e}\approx n_{0i}\approx 2
\times 10^{10}$\,cm$^{-3}$. The density profiles, $n_{e}(r)\approx
n_{i}(r)$, follow the Gaussian shape of the neutral atom cloud.

Because of the small electron-ion mass ratio, the electrons have
an initial kinetic energy approximately equal to the difference
between the photon energy and the ionization potential, typically
between 1 and $1000$\,K. The initial kinetic energy for the ions
is close to the kinetic energy of neutral atoms in the MOT. As we
will discuss below, the resulting non-equilibrium plasma evolves
rapidly.

\begin{figure}

 \mbox{\small
 \begin{minipage}[b]{5.15in}
  \begin{minipage}[t]{1in}
        \mbox{}\\ \unitlength0.75cm
           \includegraphics[width=3in,clip=true]{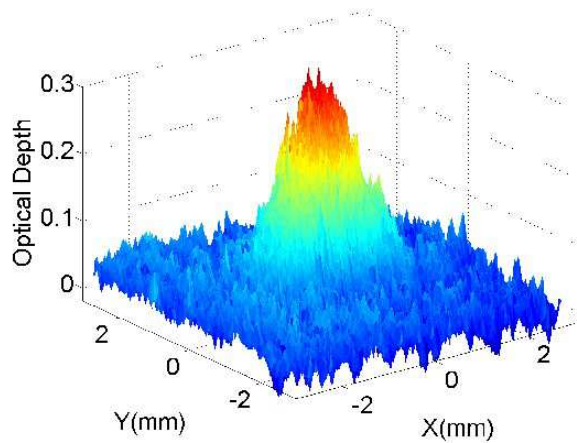}\\
  \end{minipage} \hfill
  \parbox[t]{2.8in}{\makebox[0cm]{}{\caption{Optical depth of an ultracold neutral plasma.
  The delay between the formation of the plasma and
image exposure is $85$~ns.  The plasma contains
  $7 \times 10^7$ ions and the initial %delay of 190 ns
%del190nsnd0mot06ct29494-6.000msh0usd5.000dac4atoms.dat
  peak ion density is $n_{0i}=2 \times 10^{10}$~cm$^{-3}$.
  Resolution is about $65$\,$\mu$m, limited
  by pixel averaging to improve the signal-to-noise ratio.
  }\label{image}
}}
 \end{minipage}}
\end{figure}

To record an absorption image of the plasma, a collimated laser
beam, tuned near resonance with the principle transition in the
ions, illuminates the plasma and falls on an image intensified CCD
camera. The ions scatter photons out of the laser beam and create
a shadow that is recorded by an intensified CCD camera. The
optical depth ($OD$) is defined in terms of the image intensity
without ($I_{background}$) and with ($I_{plasma}$) the plasma
present,
\begin{eqnarray}\label{ODexperiment}
OD(x,y)&=&{\rm ln}(I_{background}(x,y)/I_{plasma}(x,y)) .
\end{eqnarray}
 Figure \ref{image} shows a typical absorption
image. Section \ref{spectrumsection} describes how detailed
information about the plasma is extracted from the optical depth.

%    Spectrum of integral of optical depth over field
%        fit to a Voigt yields Doppler broadening
%        temperature is average temperature

%        It will be shown that for a plasma in which the
%        temperature varies as $T(r)=T_{max}{\rm
%        exp}(-r^2/6\sigma_i^2)$, and which expands due to the electron pressure,
%        the effective temperature extracted
%        from the Doppler broadening can be identified
%        as $T_{i,eff} \approx T_{i,ave}[1+(t/t_{exp})^2]$, where
%        $T_{i,ave}=\int_ d^3r \hspace{.025in}
%       n_i(r) T_i(r)$ is the average temperature in the plasma,
%       and $t_{exp}$ is a characteristic time scale for the
%       expansion.

\section{Ion Dynamics}
\label{plasmadynamics} In order to understand the details of the
image analysis, it is necessary to understand the dynamics of the
plasma.
 The imaging probe is most sensitive to the ion dynamics, so we will concentrate
 on this topic. The behavior of electrons was studied experimentally in
 \cite{kkb99, kkb00, klk01} and theoretically in
 \cite{kon02, mck02, rha03}.

 Ions are created with very little kinetic energy, but
 their initial spatially uncorrelated state
possesses significant Coulomb potential energy
  compared to the regular lattice that represents the
ground state of the system \cite{mbd98, mbh99}. As ions
equilibrate and correlations develop, the kinetic energy
increases. This process is called disorder-induced heating, or
correlation heating, and it has been discussed in many theoretical
papers. Early interest was generated by non-equilibrium plasmas
created by fast-pulse laser irradiation of solid targets, for
example \cite{bsk97, zwi99, mbm01, mno03}, and the problem has
been revisited in the context of ultracold neutral plasmas
\cite{kon02, mck02, mur01, ppr04jphysb}.

Qualitatively, one expects the ion temperature after equilibration
to be on the order of the Coulomb interaction energy between
neighboring ions. A quantitative analysis \cite{mur01}, assuming
complete initial disorder and incorporating the screening effects
of the
 electrons, predicts an ion temperature of
\begin{eqnarray}\label{iontemp}
  T_i&=&{2 \over 3} {e^2 \over
 4\pi \varepsilon_0 a k_B}\mid \tilde{U} +{\kappa \over
 2}\mid .
\end{eqnarray}
 Here, $\kappa =a/\lambda_D$ where
 $\lambda_D=(\varepsilon_0 k_B T_e/ n_{e} e^2)^{1/2}$ is the Debye
 length. The
 quantity $\tilde{U}\equiv {U \over N_i e^2 /
 4\pi \varepsilon_0 a} $ is the excess potential energy per particle
 in units of $e^2/4\pi \varepsilon_0 a$,
 where  $a=(4\pi n_i/3)^{-1/3}$ is the Wigner-Seitz radius, or interparticle
distance. $N_i$ is the number of  ions. $\tilde{U}$ has
 been studied with molecular dynamics simulations \cite{fha94}
 for a homogeneous system of particles interacting through a
 Yukawa
 potential, $\phi(r)= {e^2 \over 4\pi \varepsilon_0 r}{\rm
 exp}(-r/\lambda_D)$, which describes ions in the background of weakly
 coupled electrons\footnote{As the number of electrons per Debye sphere ($\kappa^{-3}$)
approaches unity, the Yukawa interaction ceases to accurately
describe ion-ion interactions. For strontium plasmas studied here,
this situation only occurs for the highest $n_e$ and lowest $T_e$.
It will be interesting to test Eq.\ \ref{iontemp} for these
conditions.}.

 For typical strontium plasmas discussed here,
 $\kappa \approx 0.1 - 1$, and $\lambda_D \approx 2 - 8$\,$\mu$m.
 $\tilde{U}$
 ranges from $-0.6$ to $-0.8$, so $T_i$ is close to
 ${e^2/
 4\pi \varepsilon_0 a k_B}$ as expected. $\kappa$ is related to the
Coulomb coupling parameter for electrons, $\Gamma_e$, through
  $\kappa=\sqrt{3\Gamma_e}$. A system of particles with charge $q$
  is strongly coupled when
  $\Gamma \equiv {q^2/
 4\pi \varepsilon_0 a k_B T}>1$ \cite{ich82}.
  $\Gamma_e\approx 0.1 - 0.5$ for the systems studied here, so the
  electrons are not strongly coupled. This
  avoids excessive complications that arise when $\Gamma_e$
approaches or initially exceeds unity, such as screening of the
ion interaction \cite{kon02}, and rapid collisional recombination
and heating of the electrons \cite{kon02,mck02,rha03,tya00},
although we do see some signs of these effects, even in this
regime. The ions typically equilibrate with $T_i\approx1$\,K,
which gives $\Gamma_i\approx 3$, so the ions are strongly coupled.

 The time scale for disorder-induced heating is the inverse of the
 ion plasma
 oscillation frequency,
 $1 /\omega_{pi}= \sqrt{m_i \varepsilon_0/n_{i}
e^2}$, which is on the order of 100 nanoseconds. Physically, this
is the time for an ion to move about an interparticle spacing when
accelerated by a typical Coulomb force of $ {e^2 /
 4\pi \varepsilon_0 a^2}$. This time scale is also evident in
 molecular dynamics
 simulations  of ion-ion
 thermalization \cite{kon02, mck02, bsk97, zwi99, mbm01, mno03, mur01,
ppr04jphysb}.

%It is interesting to note that under usual conditions in weakly
%interacting plasmas or even atomic gases, the time scale for
%relaxation of  the two-particle distribution function, which
%describes correlations, is much faster than the collision time
%that governs the relaxation of the one-particle distribution
%function to the Maxwell-Boltzmann form. This is known as
%Bogoliubov's hypothesis \cite{nic92}. For strongly-coupled
%plasmas, however, these time scales both become equal to the
%inverse of the plasma oscillation frequency \cite{zwi99}.

 %and oscillations are essentially unobservable for $\Gamma < 1$.
 %Even in an equilibrium strongly-coupled plasma, similar
 %oscillations are observed in the velocity autocorrelation
 %function \cite{hpm74,hmp75}. In a laser-produced plasma, such as those studied
 %here, oscillation of
 %the average kinetic energy is observable because the oscillatory
 %motion of each ion is synchronized by the laser pulse that
 %creates the system.

%            damping
%            electron screening

%    Expansion
%        hydrodynamic equations
%        acceleration assuming constant temperature
%            data
%        analytic solution including electron cooling
%            long time expansion

For $t_{delay}\gg 1 /\omega_{pi}$, the ions have equilibrated and
the thermal energy of the electrons begins to dominate the
evolution of the plasma. Electrons contained in the potential
created by the ions exert a  pressure on the ions that causes an
outward radial acceleration. This was studied experimentally in
\cite{kkb00} and theoretically by a variety of means in
\cite{rha03}. The experiments measured the final velocity that the
ions acquired, which was approximately $v_{terminal}\approx
\sqrt{E_e/m_i}$. With the imaging probe, we can now observe the
expansion dynamics at much earlier times during the acceleration
phase.

As discussed in  \cite{kkb00} and \cite{scg04}, a  hydrodynamic
model, which describes the plasma on length scales larger than
$\lambda_D$, shows that the pressure of the electron gas drives
the expansion through an average force per ion of
\begin{eqnarray}\label{ionforce}
\bar{F}&=& {-{\bar \nabla}(n_e(r) k_B T_e) \over n_i(r)} \approx
\hat{r} {r k_B T_e \over \sigma_i^2},
\end{eqnarray}
 where the electron and ion density distributions are
 $n_{e}(r)\approx n_{i}(r)=n_{0i}{\rm
 exp}(-r^2/2\sigma_i^2)$. We
 assume thermal equilibrium for the electrons throughout the
 cloud \cite{rha03}.

 The force leads to an average radial expansion velocity for
 the ions,
 \begin{equation}\label{ionvelocity}
    \bar{v}(r,t_{delay})=\hat{r}{r k_B T_e \over m_i\sigma_i^2}t_{delay}.
\end{equation}
 The velocity is correlated
with position and increases linearly with time.
 %$v_r(r)={r k_B T_e \over \sigma_i^2 m_i} t_{delay}$.
 This does not represent an increase in the random thermal
 velocity spread or temperature of the ions.
 Due to the large mass difference, thermalization of ions and
 electrons \cite{kon02} is slow and occurs on a millisecond time scale.
 %The increase in rms velocity due to thermalization is approximately one
 %order of magnitude
 %smaller than the increase caused by expansion.

%The rms velocity measured by the optical probe reflects the
%thermal ion motion and the radial expansion. For the purposes of
%illustrating this, we make a simplifying assumption of global
%thermal equilibrium in the plasma. (We will discuss below in Sec.
%\ref{spectrum} that this may not be a good description of the
%experiment and explain how we deal with this complication.) The
%mean squared velocity component along the imaging laser is
%\begin{eqnarray}\label{ionvz}
%&&\langle v_z^2\rangle= \int d^3r\, dv_T {n_i(r)\over N_i} P(v_T)
%(v_T+v_r(r)cos\theta )^2,
%\end{eqnarray}
% where
%\begin{equation}\label{thermalvelocitydist}
%    P(v_T)={2 \pi k_B T_e \over m_i}^{-1/2}{\rm exp}[-v_T^2/(2 {k_B T_e \over m_i})]
%\end{equation}
%is a thermal distribution of velocity along the laser direction
% for $T_i=1.4$~K.
% The behavior is
% $\lim_{t_{delay} \to 0} \sqrt{\langle v_z^2 \rangle}=\sqrt{k_B T_i \over
% m_i}$,
% and
% $\lim_{t_{delay} \to \infty} \sqrt{\langle v_z^2 \rangle}={k_B T_e \over
% m_i \sigma_i} t_{delay}$.%

%The behavior is
% $ \sqrt{\langle v_z^2 \rangle}$ is well approximated as ....

Equation \ref{ionvelocity} for the average ion velocity assumes a
constant electron temperature. Actually, as the plasma expands,
electrons will cool. This can be thought of in terms of energy
conservation or adiabatic expansion. It is possible to describe
the expansion  with a Vlasov equation  that includes the changing
electron temperature. For an initial Gaussian density
distribution, the equations can be solved analytically and the
expansion preserves the Gaussian shape with a $1/\sqrt{e}$ density
radius given by $\sigma_i^2(t) \approx \sigma_i^2(0) +[k_B
T_e(0)/m_i]t_{delay}^2$ \cite{ rha03, dse98, ppr04}. The experiments
involving absorption imaging of the plasma, however, have
concentrated on the first few microseconds of the expansion when
the plasma size and electron temperature have not changed
significantly. Thus we can safely use Eq.\ \ref{ionvelocity}. The
effects of the expansion are evident in the radial velocity that
manifests itself in Doppler broadening of the ion absorption
spectrum.

% \begin{equation}\label{expansionradius}
%    \bar{\sigma(t)}=.
%\end{equation}

%Robicheaux and Hanson \cite{rha03} and Pohl, Pattard, and Rost
%\cite{ppr04} performed a more complete analysis that implied
%\begin{equation}\label{expansion}
%    \sigma_i(t)^2=\sigma_i(0)^2+{k_B T_e(0) \over m_i}t^2.
%\end{equation}
%The functional form agrees with what we observe in our data (Fig.\
%\ref{paperalltemp}). The expression implies an expansion velocity
%$v_{0}=\sqrt{{k_B T_e(0) \over m_i}}$.

%Optical probe measures early times

%%Previous experiment measures terminal velocity measured terminal
%velocity, which agrees with Eq.\ \ref{expansionradius}.

%We can also make use of the images to directly see the expanding
%density profile. This will only be visible on a longer time scale,
%$\sigma_i/v_{terminal}$ and we have not yet pursued these studies.

\section{Doppler-Broadened Spectrum}
\label{spectrumsection}

\begin{figure}
 \mbox{\small
 \begin{minipage}[b]{5.15in}
  \begin{minipage}[t]{1in}
        \mbox{}\\ \unitlength0.75cm
           \includegraphics[width=3in,clip=true]{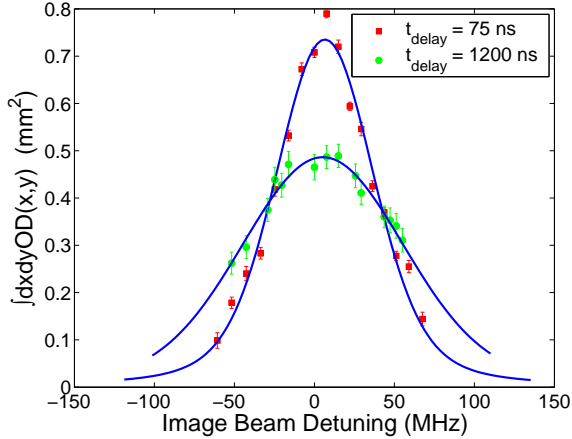}\\
  \end{minipage} \hfill
  \raisebox{.15in}{\parbox[t]{3in}{\makebox[0cm]{}{\caption{Absorption spectra of ultracold neutral plasmas.
  We plot the integral of the optical depth (Eq.\ \ref{ODintegral}).
  The frequency is with respect to a Doppler-free absorption
  feature in a strontium discharge cell.
  Both spectra correspond to $T_e=56$~K and
  an initial peak plasma density of
  $n_{0i}=2 \times 10^{10}$~cm$^{-3}$.
  Data are fit with Voigt profiles, and the increase in
  linewidth for longer $t_{delay}$ is clear.}
  \label{spectrum}
}}}
 \end{minipage}}
\end{figure}

To  obtain quantitative information from the plasma images, we
relate the $OD$ (Eq.\ \ref{ODexperiment}) to underlying physical
parameters. Following Beer's law, the $OD$ for a laser propagating
along the z axis is
\begin{eqnarray}\label{ODtheory}
OD(x,y)&=&\int dz \hspace{.025in}
       n_i(x,y,z) \alpha[\nu, T_i(r)],
\end{eqnarray}
 where
 $n_{i}(x,y,z)$ is the ion density, and $\alpha[\nu, T_i(r)]$ is the
 ion absorption cross section at
 the image beam
frequency, $\nu$. The  absorption cross section is a function of
temperature due to Doppler broadening, and since we expect the
temperature to vary with density, we allow $\alpha$ to vary with
position. If we now integrate over x and y, or, in reality, sum
over the image pixels multiplied by the pixel area, we get the
spectrum
\begin{eqnarray}\label{ODintegral}
S(\nu)\equiv \int dx dy OD(x,y)&=&\int d^3r \hspace{.025in}
       n_i(r) \alpha[\nu, T_i(r)],
\end{eqnarray}
as a function of the image laser detuning\footnote{We can also
fit $OD(x,y)$ to a two dimensional Gaussian, as described in
\cite{scg04}, and identify  $\int dx dy OD(x,y)\approx 2\pi
\sigma_{ix}\sigma_{iy} OD_{max}$, where $\sigma_{ix}$ and
$\sigma_{iy}$ are the transverse sizes of the absorption profile,
and $OD_{max}$ is the peak optical depth. This sometimes has
signal-to-noise ratio advantages over integrating the entire
image, but both approaches should give the same result.}.  As we
vary the detuning, we obtain absorption spectra as shown in Fig.\
\ref{spectrum}. The rest of the paper will deal with the
relationship between spectra such as these and the underlying
temperature distributions of the ions.

 The
absorption cross section for ions in a region described by a
temperature $T_i$, is given by the Voigt profile
\begin{eqnarray} \label{absorptioncrossection}
  \alpha(\nu, T_i)&=&\int d s {3^*\pi \lambda^2 \over
  2}{\gamma_0 \over \gamma_{eff}}{1 \over 1+ 4( { \nu-s \over \gamma_{eff}/2\pi} )^2} {1
  \over \sqrt{2\pi} \sigma_D(T_i)} {\rm e}^{-(s-\nu_0)^2/2\sigma_D(T_i)^2}, \nonumber \\
  \end{eqnarray}
where $\sigma_D(T_i)=\sqrt{k_B T_{i}/m_i}/\lambda$ is the Doppler
width, and $\gamma_{eff}=\gamma_0+\gamma_{laser}$ is the effective
Lorentizian linewidth due to the natural linewidth of the
transition, $\gamma_0=2\pi \times 22 \times 10^6$\, rad/s, and the
laser linewidth, $\gamma_{laser}=2\pi \times (5 \pm 2)\times
10^6$\, rad/s. The center frequency of the transition is
$\nu_0=c/\lambda$, where $\lambda=422$\,nm. The ``three-star"
symbol, $3^*=1$, accounts for the equal distribution of ions in
the doubly degenerate ground state and the linear polarization of
the imaging light \cite{sie86}.
%We have used Beer's law to relate $S(\nu)$ to the absorption cross
%section,  $\alpha[\nu,T_{i}(r)]$, and the ion density. The
%absorption cross section is a function of temperature due to the
%Doppler broadening, and since we expect the temperature to vary
%with density, we allow $\alpha$ to vary with position. The
%absorption cross section has a Voigt profile resulting from the
%convolution of the Gaussian Doppler broadening with a Lorentzian
%of width $\gamma=\gamma_0+\gamma_{laser}$. Here,
%$\gamma_0=22$\,MHz is the natural width of the transition, and
%$\gamma_{laser}=(10 \pm 2)$\,MHz is the measured linewidth of the
%laser.
%As shown in the second line of Eq.\ \ref{OD},
We fit the spectrum to
a single Voigt profile with effective temperature
 $T_{i,eff}$,
 \begin{eqnarray}\label{OD}
S(\nu)&=&\int
\mathrm{d}^{3}r
 \hspace{0.1cm}{n_{i}(r)}
\alpha[\nu,T_{i}(r)] \equiv N_{i}\alpha(\nu,T_{i,eff}),
\end{eqnarray}
where $N_{i}$ is the number of ions.
 We will explain below that $T_{i,eff}$ is a good approximation
 of the  average ion temperature in the plasma.
 %Plotting $T_{i,eff}$ versus
 %delay time thus probes the evolution of the ion kinetic energy.
  %$T_{C}=e^2/4\pi \varepsilon_0 a k_B$

\section{Recent Results}

\label{resultssection}

\begin{figure}
%print to pdfwriter from ppoint,open resulting pdf with ghostview, convert to
%eps with pswrite, max resoltuion (by this I mean use convert, select pswrite with max res and put an .eps extension on)
  %\includegraphics[width=4in,clip=true,trim=100 290 50 80 ]{fig1.eps}
  \includegraphics[width=4in,clip=true]{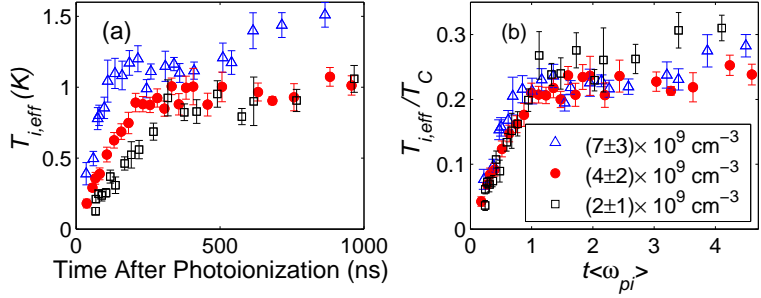}\\
  \caption{The effective ion temperature, $T_{i,eff}$, versus time
  after photoionization for
  initial
electron temperature of $T_e=2 E_e/3k_B=38\pm 6$\,K and
  various plasma densities. (A) The data is plotted on absolute temperature and
  time scales. (B) The
  time is scaled by the inverse of the average plasma period, and
  $T_{i,eff}$ is
  scaled by $T_{C}$. From \cite{csg04}.
}
 \label{fig1}
\end{figure}

Fig \ref{fig1}(a) shows the evolution of  $T_{i,eff}$ for
three different densities. The rapid increase in the temperature
for $t_{delay}<300$ ns is due to disorder-induced heating. This
originates from conversion of Coluomb potential energy into
kinetic energy as the ions evolve from a completely disordered
state to one with some degree of spatial correlations. This was
predicted in \cite{mur01}, and observed in numerical simulations
\cite{kon02,tya00,ppr04jphysb} and the first experimental studies
with optical imaging \cite{scg04}.

The time scale of the heating is the inverse ion plasma frequency
$\omega_{pi}^{-1}=\sqrt{m_i \varepsilon_0/n_{i} e^2}\approx
100$\,ns. The energy scale is $T_C=e^2/4\pi \varepsilon_0 a
k_B\approx 5$\,K, where $a=(4\pi n_{i}/3)^{-1/3}$ is the
Wigner-Seitz radius. Fig \ref{fig1}(b) shows the same data with time
scaled by the average inverse plasma frequency and temperature
scaled by $T_C$. The three curves coincide very well in the time
axis, but show slight deviation in the temperature axis. The
deviation indicates the effects of electron screening of the
ion-ion interaction. Electron screening is discussed
in \cite{csg04}.

To quantitatively analyze the data in Figs.\ \ref{fig1}
 we must account for
 the effect of the inhomogeneous density
distribution
\cite{kcg04archive}.
We expect the ion temperature to vary with density
because global thermal equilibrium occurs on a hydrodynamic time
scale, $\sigma/v$, which is on the order of tens of $\mu$s, where
$v$ is the ion acoustic wave velocity. Local thermal equilibration
occurs on a much faster time scale, $\sim \omega_{pi}^{-1}$
\cite{ppr04archive}. For a range of reasonable temperature
distribution,  numerical simulations \cite{kcg04archive}
show that $T_{i,eff}=(0.95\pm0.05) T_{i,ave}$. Here $T_{i,ave}$ is
the average ion temperature in the plasma. The extracted $T_{i,ave}$
and Eq.\ \ref{iontemp} agree  within our uncertainties of typically
0.2\,K.

If the effective ion temperature is measured for longer delay times,
acceleration of the ions by the electron pressure also becomes important.
The effect is small for the data presented here, but its effect can be
modelled \cite{csg04,kcg04archive}. The acceleration of the ions then
becomes a sensitive probe of the electron temperature which
evolves under the influence of recombination, continuum lowering,
disorder-induced heating, and adiabatic expansion
 \cite{kkb00,scg04,csg04,rha03,ppr04}.

From the measured $T_{i,ave}$, $T_e$, and density distribution,
%Wigner-Seitz radius,
% $a_{ave}=\langle(4\pi n_{0i}/3)^{-1/3}\rangle$,
we calculate the average Coulomb coupling constant for
Debye-screend ions,
\begin{equation}\label{gammaave}
    \Gamma_{i,avg}^*= \langle{\rm
exp}(-\kappa(r))e^2 / (4\pi \varepsilon_0 a(r) k_B
T_{i,ave})\rangle.
\end{equation}
%But for Debye
%screened particles, the Coulomb coupling constant is reduced to
%$\Gamma_{i}^{*}=\Gamma_{i}\exp^{-\kappa}$ due to screening.
For all the data shown in Figs. \ref{fig1}  and in
\cite{csg04},
$\Gamma_{i,avg}^{*}$ is in the range of 1.7 to 2.5. With lower
$T_{e}$ and higher density, $\Gamma_{i,ave}^{*}$ is slightly
higher. The surprisingly small variation in $\Gamma_{i,avg}^*$
suggests that disorder-induced heating is a natural feedback
mechanism that leads to equilibration just barely in the strongly
coupled regime.

\begin{figure}
  \includegraphics[width=4in,clip=true]{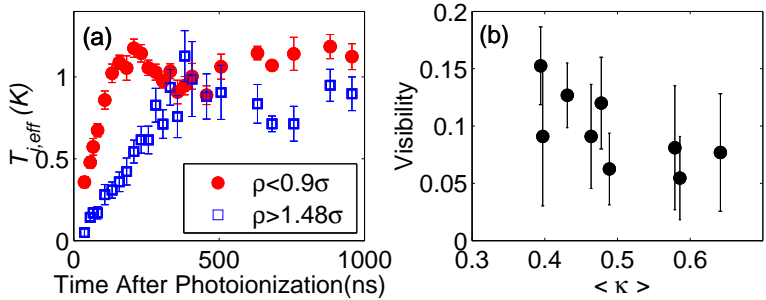}\\
  \caption{Effective ion temperature obtained from different
  selected regions of the cloud for
  $n_{0i}=(4\pm 2)\times 10^{9}$\,cm$^{-3}$
  and initial $T_e=2 E_e/3k_B=38\pm 6$\,K.
   In (b), we plot the visibility of
  the oscillation for the central probed region versus the averaged
  $\kappa$ for this region for all our data
  taken at different conditions. From \cite{csg04}.
  }\label{fig4}
\end{figure}

Close inspection of Figs.\ \ref{fig1}
reveals that at the end of the disorder-induced heating phase, the
ion temperature overshoots it equilibrium value before settling.
This phenomenon is more evident in Fig.\ \ref{fig4}(a), where
$T_{i,ave}$ is calculated for an inner and outer region of the
plasma image ($\rho=\sqrt{x^2+y^2}< 0.9 \sigma$ and $\rho>1.48
\sigma$ respectively). Each selected annular region contains 1/3
of the ions and probes a region with significantly less variation
in density than in the entire plasma. The region with lower
density has lower $T_{i,ave}$, supporting our hypothesis of local
thermal equilibrium, but the oscillation is the most striking
observation.

Intuitively, one can explain this phenomenon as an oscillation of
each ion in its local potential energy well. A simple calculation
implies that the time for an ion to move an interparticle
distance, when accelerated from rest by a force of $e^2/(4\pi
\varepsilon_0 a^2)$, varies as $\sim \omega_{pi}^{-1}$, and the
observed  oscillation occurs at $2 \omega_{pi}$. As expected from
the density dependence of $\omega_{pi}$, the oscillation period is
longer for the outer region where the average density is lower.
This explains why averaging over the entire cloud obscures the
oscillation; the motion dephases because of the variation in
$\omega_{pi}$. It is questionable whether the ion motion should be
called an ion plasma oscillation or not, because there is probably
no collective, or long range coherence to the motion.

Kinetic energy oscillations at $2 \omega_{pi}$ have been observed
in molecular dynamics simulations of equilibrating
strongly-coupled systems \cite{zwi99,mno03,ppr04jphysb}.
Calculations \cite{hpm74,hmp75} also show oscillations in the
velocity autocorrelation function in equilibrium strongly-coupled
one-component plasmas. To our knowledge, this is the first
experimental observation of the phenomenon.
%In a laser-produced plasma, such as those studied
% here, oscillation of
% the average kinetic energy is observable because the oscillatory
% motion of each ion is synchronized by the laser pulse that
% creates the system.

Numerical results\cite{zwi99} suggest that the damping time for
the oscillations is approximately $ \pi / \omega_{pi}$ for
$\Gamma_i \ge 5$, and that lower $\Gamma_i$ leads to faster
damping. Because our analysis still averages over the z-axis of
the plasma, which introduces dephasing into the observed
oscillations, it is difficult to comment on the damping of the
oscillation. But we do find a correlation between the visibility of the
oscillation and $\kappa$, as shown in Fig.\ \ref{fig4}(b), suggesting that
electron screening plays a role in the damping. The
visibility is defined as
$(T_{i,eff}^{peak}-T_{i,eff}^{dip})/(T_{i,eff}^{peak}+T_{i,eff}^{dip})$,
where $T_{i,eff}^{peak}$ and $T_{i,eff}^{dip}$ are effective ion
temperatures at the peak and dip of the oscillation.

In conclusion, we have developed absorption imaging as a tool
for studying ultracold neutral plasmas
\cite{scg04}.
The absorption spectrum, after accounting for
the inhomogeneous plasma density distribution, measures the
average ion temperature.
Equilibration of the plasma shows the effect of
disorder-induced heating.
The screening effect of electrons on the final
ion temperature has been studied, and experiment and theory agree very
well \cite{csg04}.
The ion kinetic energy also displays oscillations that
are characteristic of equilibrating strongly coupled plasmas.

Many future experiments suggest themselves. Some of the most
interesting are investigating dynamics when the initial electron
Coulomb coupling parameter is large and recombination and
disorder-induced electron heating  are expected to dominate the
plasma evolution. Detailed examination of ion and electron
thermalization at the border of the strongly coupled regime is
also possible. Quantitative study of the
kinetic energy oscillations is also necessary in order
to develop a better understanding  of aspects of the phenomenon
such as damping.
Improvements in imaging optics will also increase the
image signal-to-noise ratio and allow study of features on the ion
density distribution with $\sim 10$~$\mu$m experimental
resolution.

The fact that strontium ions scatter blue light so efficiently suggests
one of the most exciting
future directions for studies of ultracold nuetral plasmas.
It may
be possible to use laser light to cool and trap the ions in a
strontium ultracold neutral plasma  \cite{kag03}, just as lasers are used to
cool and trap neutral atoms. Confining the ions would also confine
the electrons through Coulomb attraction. This would be an
entirely new method of plasma confinement and may also lead to
drastically lower plasma temperatures and stronger coupling for the
ions.

This research was supported by the Department of Energy Office of
Fusion Energy Sciences, National Science Foundation, Office for
Naval Research, Research Corporation, Alfred P. Sloan Foundation,
and David and Lucille Packard Foundation.
\\

%\bibliography{bibliography}

\begin{thebibliography}{10}

\bibitem{kkb99}
T.~C. Killian, S.~Kulin, S.~D. Bergeson, L.~A. Orozco, C.~Orzel, and S.~L.
  Rolston.
\newblock Creation of an ultracold neutral plasma.
\newblock {\em { Phys. Rev. Lett.}}, 83(23):4776, 1999.

\bibitem{ich82}
S.~Ichamuru.
\newblock Strongly coupled plasmas: high-density classical plasmas and
  degenerate electron liquids.
\newblock {\em { Rev. Mod. Phys.}}, 54(4):1017, 1982.

\bibitem{vho91}
H.~M.~Van Horn.
\newblock Dense astrophysical plasmas.
\newblock {\em { Science}}, 252:384, 1991.

\bibitem{nmg98}
M.~Nantel, G.~Ma, S.~Gu, C.~Y. Cote, J.~Itatani, and D.~Umstadter.
\newblock Pressure ionization and line merging in strongly coupled plasmas
  produced by 100-fs laser pulses.
\newblock {\em { Phys. Rev. Lett.}}, 80(20):4442, 1998.

\bibitem{sht00}
E.~Springate, N.~Hay, J.~W.~G. Tisch, M.~B. Mason, T.~Ditmire, M.~H.~R.
  Hutchinson, and J.~P. Marangos.
\newblock Explosion of atomic clusters irradiated by high-intensity laser
  pulses: Scaling of ion energies with cluster and laser parameters.
\newblock {\em { Phys. Rev. A}}, 61:063201, 2000.

\bibitem{mtk99}
G.~E. Morfill, H.~M. Thomas, U.~Konopka, and M.~Zuzic.
\newblock The plasma condensation: Liquid and crystalline plasmas.
\newblock {\em { Phys. Plasmas}}, 6(5):1769, 1999.

\bibitem{mbh99}
T.~B. Mitchell, J.~J. Bollinger, X.~{-P.} Huang, W.~M. Itano, and D.~H.~E.
  Dubin.
\newblock Direct observations of the structural phases of crystallized ion
  plasmas.
\newblock {\em { Phys. Plasmas}}, 6(5):1751, 1999.

\bibitem{kkb00}
S.~Kulin, T.~C. Killian, S.~D. Bergeson, and S.~L. Rolston.
\newblock Plasma oscillations and expansion of an ultracold neutral plasma.
\newblock {\em { Phys. Rev. Lett.}}, 85(2):318, 2000.

\bibitem{klk01}
T.~C. Killian, M.~J. Lim, S.~Kulin, R.~Dumke, S.~D. Bergeson, and S.~L.
  Rolston.
\newblock Formation of {R}ydberg atoms in an expanding ultracold neutral
  plasma.
\newblock {\em { Phys. Rev. Lett.}}, 86(17):3759, 2001.

\bibitem{rtn00}
M.~P. Robinson, B.~L. Tolra, M.~W. Noel, T.~F. Gallagher, and P.~Pillet.
\newblock Spontaneous evolution of rydberg atoms into an ultracold plasma.
\newblock {\em { Phys. Rev. Lett.}}, 85(21):4466, 2000.

\bibitem{scg04}
C.~E. Simien, Y.~C. Chen, P.~Gupta, S.~Laha, Y.~N. Martinez, P.~G. Mickelson,
  S.~B. Nagel, , and T.~C. Killian.
\newblock Using absorption imaging to study ion dynamics in an ultracold
  neutral plasma.
\newblock {\em {Phys. Rev. Lett.}}, 92(14):143001, 2004.

\bibitem{kag03}
T.~C. Killian, V.~S. Ashoka, P.~Gupta, S.~Laha, S.~B. Nagel, C.~E. Simien,
  S.~Kulin, S.~L. Rolston, and S.~D. Bergeson.
\newblock Ultracold neutral plasmas: recent experiments and new prosects.
\newblock {\em { J. Phys. A: Math. Gen.}}, 36:6077, 2003.

\bibitem{nsl03}
S.~B. Nagel, C.~E. Simien, S.~Laha, P.~Gupta, V.~S. Ashoka, and T.~C. Killian.
\newblock Magnetic trapping of metastable {$^3P_2$} atomic strontium.
\newblock {\em { Phys. Rev. A}}, 67:011401, 2003.

\bibitem{mvs99}
H.~J. Metcalf and P.~van~der Straten.
\newblock {\em Laser Cooling and Trapping}.
\newblock Springer-Verlag New York, New York, 1999.

\bibitem{kon02}
S.~G. Kuzmin and T.~M. O'Neil.
\newblock Numerical simulation of ultracold plasmas.
\newblock {\em { Phys. Plasmas}}, 9(9):3743, 2002.

\bibitem{mck02}
S.~Mazevet, L.~A. Collins, and J.~D. Kress.
\newblock Evolution of ultracold neutral plasmas.
\newblock {\em { Phys. Rev. Lett.}}, 88(5):55001, 2002.

\bibitem{rha03}
F.~Robicheaux and J.~D. Hanson.
\newblock Simulated expansion of an ultra-cold, neutral plasma.
\newblock {\em { Phys. Plasmas}}, 10(6):2217, 2003.

\bibitem{mbd98}
T.~B. Mitchell, J.~J. Bollinger, D.~H.~E. Dubin, X.~{-P.} Huang, W.~M. Itano,
  and R.~H. Baughman.
\newblock Direct observations of structural phase transitions in planar
  crystallized ion plasmas.
\newblock {\em { Science}}, 282:1290, 1998.

\bibitem{bsk97}
M.~Bonitz, D.~Semkat, and D.~Kremp.
\newblock Short-time dynamics of correlated many-particle systems: Molecular
  dynamics versus quantum kinetics.
\newblock {\em { Phys. Rev. E}}, 56(1):1246, 1997.

\bibitem{zwi99}
G.~Zwicknagel.
\newblock Molecular dynamics simulations of the dynamics of correlations and
  relaxation in an ocp.
\newblock {\em { Contrib. Plasma Phys.}}, 39:155, 1999.

\bibitem{mbm01}
K.~Morawetz, M.~Bonitz, V.~G. Morozov, G.~R{\"o}pke, and D.~Kremp.
\newblock Short-time dynamics with initial correlations.
\newblock {\em { Phys. Rev. E}}, 63:020102, 2001.

\bibitem{mno03}
I.~V. Morozov and G.~E. Norman.
\newblock Non-exponential dynamic relaxation in strongly nonequilibrium
  nonideal plasmas.
\newblock {\em { J. Phys. A: Mah. Gen.}}, 36:6005, 2003.

\bibitem{mur01}
M.~S. Murillo.
\newblock Using fermi statistics to create strongly coupled ion plasmas in atom
  traps.
\newblock {\em { Phys. Rev. Lett.}}, 87(11):115003, 2001.

\bibitem{ppr04jphysb}
T.~Pohl, T.~Pattard, and J.~M. Rost.
\newblock On the possibility of `correlation cooling' of ultracold neutral
  plasmas.
\newblock {\em { J. Phys. B: At. Mol. Opt. Phys.}}, 37:183, 2004.

\bibitem{fha94}
R.~T. Farouki and S.~Hamaguchi.
\newblock Thermodynamics of strongly-coupled {Y}ukawa systems near the
  one-component-plasma limit. {II}. {M}olecular dynamics simultations.
\newblock {\em { J. Chem. Phys.}}, 101(11):9885, 1994.

\bibitem{tya00}
A.~N. Tkachev and S.~I. Yakovlenko.
\newblock Moderation of recombination in an ultracold laser-produced plasma.
\newblock {\em { Quantum Electronics}}, 30(12):1077, 2000.

\bibitem{dse98}
D.~S. Dorozhkina and V.~E. Semenov.
\newblock Exact solution of vlasov equations for quasineutral expansion of
  plasma bunch into vacuum.
\newblock {\em { Phys. Rev. Lett.}}, 81(13):2691, 1998.

\bibitem{ppr04}
T.~Pohl, T.~Pattard, and J.~M. Rost.
\newblock Coulomb crystallization in expanding laser-cooled neutral plasmas.
\newblock {\em { Phys. Rev. Lett.}}, 92(15):155003, 2004.

\bibitem{sie86}
A.~E. Siegman.
\newblock {\em Lasers}.
\newblock University Science Books, Sausolito, California, 1986.

\bibitem{csg04}
Y. C. Chen, C. E. Simien, P. Gupta, S. Laha, Y. N. Martinez, P. G. Mickelson,
  S. B. Nagel, and T. C. Killian. Submitted {\it Phys. Rev. Lett.}

\bibitem{kcg04archive}
T. C. Killian, Y. C. Chen, P. Gupta, S. Laha, Y. N. Martinez, P. G. Mickelson,
  S. B. Nagel , A. D. Saenz, and C. E. Simien. Submitted {\it J. Phys. B.},
  arXiv/physics/0407138.

\bibitem{ppr04archive}
T. Pohl, T. Pattard, and J. M. Rost, arXiv/physics/0405125.

\bibitem{hpm74}
J.~P. Hansen, E.~L. Pollock, and I.~R. McDonald.
\newblock Velocity autocorrelation funtion and dynamical structure factor of
  the classical one-component plasma.
\newblock {\em { Phys. Rev. Lett}}, 32(6):277, 1974.

\bibitem{hmp75}
J.~P. Hansen, I.~R. McDonald, and E.~L. Pollock.
\newblock Statistical mechanics of dense ionized matter. {III}. {D}ynamical
  properties of the classical one-component plasma.
\newblock {\em { Phys. Rev. A}}, 11(3):1025, 1975.

\end{thebibliography}

%\begin{thebibliography}{9}
%\bibitem{c1}
%A. Einstein, Phys. Rev. A {\bf 00}, 0101 (1911).
%\bibitem{c2} S. Timoshenko and S. Woinowsky-Krieger, {\it Theory of
%Plates and Shells}, McGraw-Hill, New York (1959).
%\end{thebibliography}

\end{document}